\documentclass[a4paper,twocolumn,showpacs,prb]{revtex4}
\usepackage[english]{babel}
\usepackage{graphicx}
\usepackage{xspace}

\begin{document}

\title{Hall effect, magnetization and conductivity of Fe$_3$O$_4$ epitaxial thin films}

\author{D.~Reisinger}
\email{Daniel.Reisinger@wmi.badw.de}
\affiliation{Walther-Mei{\ss}ner-Institut, Bayerische Akademie der
Wissenschaften, Walther-Mei{\ss}ner Str. 8, 85748 Garching,
Germany}

\author{P.~Majewski}
\affiliation{Walther-Mei{\ss}ner-Institut, Bayerische Akademie der
Wissenschaften, Walther-Mei{\ss}ner Str. 8, 85748 Garching,
Germany}

\author{M.~Opel}
\affiliation{Walther-Mei{\ss}ner-Institut, Bayerische Akademie der
Wissenschaften, Walther-Mei{\ss}ner Str. 8, 85748 Garching, Germany}

\author{L.~Alff}
\affiliation{Walther-Mei{\ss}ner-Institut, Bayerische Akademie der Wissenschaften,
Walther-Mei{\ss}ner Str. 8, 85748 Garching, Germany}

\author{R.~Gross}
\affiliation{Walther-Mei{\ss}ner-Institut, Bayerische Akademie der
Wissenschaften, Walther-Mei{\ss}ner Str. 8, 85748 Garching, Germany}

\date{received April 5, 2004}
\pacs{%
72.15.-v %Electronic conduction in metals and alloys
72.80.Ga %Transition-metal compounds
75.70.-i % Magnetic films and multilayers}
}

%%%%%%%%%%%%
% ABSTRACT %
%%%%%%%%%%%%

\begin{abstract}
Magnetite epitaxial thin films have been prepared by pulsed laser
deposition on MgO and Si substrates. The magnetic and electrical
properties of these epitaxial films are close to those of single
crystals. For 40 - 50\,nm thick films, the saturation
magnetization and electrical conductivity are $\sim
450$\,emu/cm$^3$ and 225\,$\Omega^{-1}\text{cm}^{-1}$ at room
temperature, respectively. The Verwey transition temperature is
117\,K. The Hall effect data yield an electron concentration
corresponding to 0.22 electrons per formula unit at room
temperature. Both normal and anomalous Hall effect have been found
to have negative sign.
\end{abstract}

\maketitle

Half-metallic materials with 100\% spin polarization of the charge carriers at
the Fermi-level are under intensive investigation due to their potential
application in spintronics. Promising candidates with theoretically predicted
half-metallicity are Fe$_3$O$_4$\cite{Zhang:91}, CrO$_2$\cite{Korotin:98}, Mn
based Heusler alloys\cite{deGroot:83}, doped manganites\cite{Pickett:96}, and
double perovskites\cite{Kobayashi:98,Philipp:03} with magnetite having the
highest Curie temperature ($T_C \simeq 850$\,K). Also, magnetite has a low
magnetic-crystalline anisotropy. For the use of magnetite in spintronic
devices, the growth of high quality thin films is required. Recently,
Fe$_3$O$_4$ thin films have been grown by different techniques including
sputtering\cite{Margulies:96}, molecular beam epitaxy\cite{Voogt:98} and pulsed
laser deposition\cite{Gong:97,Ogale:98} on various substrates (MgO,
MgAl$_2$O$_4$, SrTiO$_3$, sapphire, and Si). On Si, epitaxial thin films can be
grown despite the large lattice mismatch using a suitable buffer layer system,
for example a combination of TiN and MgO\cite{Reisinger:03a}. In this letter,
we report electrical transport, magnetization, and Hall effect measurements on
high quality epitaxial magnetite thin films grown on MgO and Si substrates by
laser molecular beam epitaxy\cite{Gross:2000a}. The key result is, that for the
high quality epitaxial films about the same properties can be achieved as for
bulk samples. In particular, we report on Hall effect measurements showing that
the magnetite thin films have a similar charge carrier concentration at room
temperature as single crystals\cite{Lavine:59}. To the best of our knowledge,
there is at present no other Hall measurement available for magnetite thin
films.

The typically 40 to 50\,nm thick manganite thin films were deposited from a
stoichiometric target by laser molecular beam epitaxy\cite{Gross:2000a} at a
substrate temperature of 340$^{\circ}$C on MgO(001) or Si(001) substrates.
In-situ RHEED was used to monitor the block-by-block growth
mode\cite{Gross:2000a,Klein:99} of magnetite. Note that for Fe$_3$O$_4$ {\em
four} RHEED intensity oscillations are observed per unit
cell\cite{Reisinger:03b} corresponding to four charge neutral blocks of
composition Fe(A)$_2^{3+}$Fe(B)$_2^{3+}$Fe(B)$_2^{2+}$O$_8^{2-}$. Here, A and B
refer to the tetrahedral and octahedral sites of the inverse spinel structure,
respectively, where the B-site is equally occupied by Fe$^{3+}$ and Fe$^{2+}$
ions. A more detailed discussion of the thin film growth process on Si is given
in Ref.~\cite{Reisinger:03a}. In order to prevent surface oxidation, the
magnetite films were covered by an about 10\,nm thick MgO cap layer. X-ray
analysis gives a Fe$_3$O$_4$ $c$-axis value of 8.39\,{\AA} and a typical mosaic
spread of $0.02^\circ$ for MgO substrates. No impurity phases, for example from
other iron oxides, could be observed in the diffraction pattern.  The rms
surface roughness as determined by AFM had typical values ranging between 2 and
5\,{\AA} averaged over an area of 1\,$\mu$m$^2$. The film thickness was evaluated
by counting the RHEED intensity oscillations and verified by x-ray
reflectometry.

\begin{figure}[tb]
\centering{%
\includegraphics[width=0.99\columnwidth,trim=10 30 10 10]{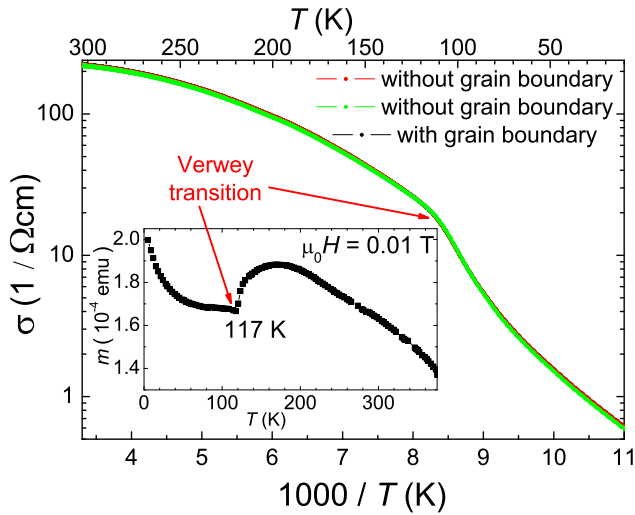}}
\vspace*{0mm}\\
\caption{Electrical conductivity of a 45\,nm thick epitaxial
Fe$_3$O$_4$ films grown on a MgO bicrystal plotted versus $1/T$.
The curves for bridges with and without grain boundary are
indistinguishable. The Verwey transition around 120\,K can be best
seen in the inset, where magnetic moment measured in a field of
100\,Oe is plotted vs. temperature. }
 \label{Reisinger_Fig1}
\end{figure}

In Fig.~\ref{Reisinger_Fig1}, the electrical conductivity $\sigma$
of an epitaxial Fe$_3$O$_4$ thin film grown on a MgO(001)
symmetrically [001] tilt bicrystal substrate with a misorientation
angle of $24.8^\circ$ degree is plotted versus $1/T$.  From this
data we can conclude the following: First, it is evident that the
$\sigma (T)$ curve obtained for a bridge straddling the grain
boundary is the same as for bridges containing no grain boundary.
That is, obviously the grain boundary resistance is much smaller
than the resistance of the adjacent film part of the bridge. A
similar effect has been observed for grain boundaries in doped
manganites, where only a high-temperature annealing process in
oxygen atmosphere produces a grain boundary with sufficiently high
resistance\cite{Philipp:00}. Second, the absolute value of the
conductivity of about $225\,\Omega^{-1}\text{cm}^{-1}$ at room
temperature is comparable to the best values of about
$250\,\Omega^{-1}\text{cm}^{-1}$ reported for bulk
samples\cite{Calhoun:54} and of about
$190\,\Omega^{-1}\text{cm}^{-1}$ reported for films of the same
(50\,nm) thickness\cite{Eerenstein:02}. Recently, a decrease of
$\sigma$ has been reported for films thinner than
50\,nm\cite{Eerenstein:02} due to a decrease of the antiphase
domain size. However, there are also reports that the conductivity
does not change down to 10\,nm\cite{Soeya:02}. Third, the Verwey
transition at $T_{\text{V}}\simeq 120$\,K appears smeared out in
the $\sigma(T)$ curve as compared to single crystals, where a
sharp jump in $\sigma (T)$ is observed. We note that also the
magnetic moment (see inset in Fig.~\ref{Reisinger_Fig1}) gradually
decreases starting already above 150\,K. This is again in contrast
to bulk samples, where a sharp jump is observed at $T_{\text{V}}$.
However, the data clearly shows a sharp kink structure at
$T_{\text{V}}=117$\,K. The physics of this Verwey transition and
its relation to a possible charge order is still under discussion.

%%%%%%%%%%%%%%%%%%%%% FIGURE 1: Conductivity %%%%%%%%%%%%%%%%%%%%%%%%%%
%%%%%%%%%%%%%%%%%%%%%%%%%%%%%%%%%%%%%%%%%%%%%%%%%%%%%%%%%%%%%%%%%%%%%%%

\begin{figure}[tb]
\centering{%
\includegraphics
[width=0.99\columnwidth]{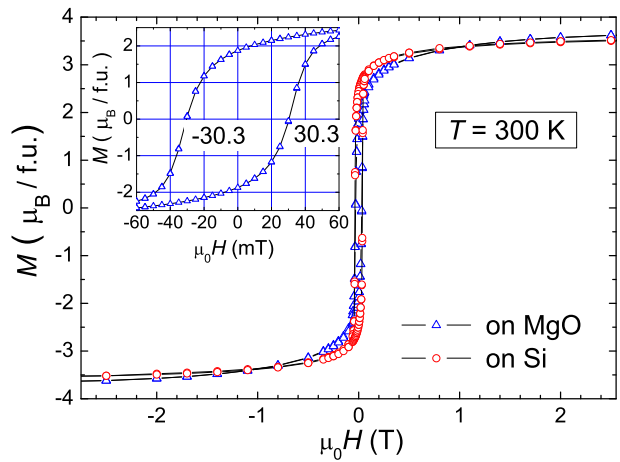}}
 \caption{Magnetization versus applied magnetic field at 300\,K for epitaxial Fe$_3$O$_4$ films
grown on MgO and Si substrates. The field was applied in-plane. The inset shows
a blowup of the low field part of the hysteresis curve for the film grown on
MgO.}
 \label{Reisinger_Fig2}
\end{figure}

Fig.~\ref{Reisinger_Fig2} shows the magnetization vs applied field dependencies
for Fe$_3$O$_4$ films grown on MgO and Si\cite{Reisinger:03a}. At 300\,K, the
saturation magnetization $M_s$ is about $3.6$ and $3.5\,\mu_B/\text{f.u.}$ for
the film on MgO and Si, respectively. At 150\,K, this value increases to
$3.8\,\mu_B/\text{f.u.}$. The coercive field is about 30\,mT at room
temperature. The magnetization data show that the $M_s$ values of our epitaxial
films with a thickness ranging between 40 and 50\,nm are close to the
theoretically expected bulk value of $M_s = 4\,\mu_B/\text{f.u.}$. Our $M_s$
values are among the highest reported for thin films in the literature so
far\cite{Margulies:96,Kale:01,Soeya:02}. We also note that our room temperature
$M_s$ value of 453\,emu/cm$^3$ is close to the value of 471\,emu/cm$^3$
reported for a single crystal\cite{LB}, i.e.~for our thin films about 96\% of
the bulk value is reached.  We attribute the high $M_s$ values of our films
their good structural properties of the films and the fact that surface
oxidation is prevented using a MgO cap layer. Due to the good lattice match
between the film and the substrate (for the Si substrate the lattice mismatch
is relaxed using a TiN/MgO double buffer layer system \cite{Reisinger:03a})
also strain effects play no important role.

%%%%%%%%%%%%%%%%%%%%% FIGURE 2: Saturation Magnetization %%%%%%%%%%%%%%%%%%%%%%%%%%
%%%%%%%%%%%%%%%%%%%%%%%%%%%%%%%%%%%%%%%%%%%%%%%%%%%%%%%%%%%%%%%%%%%%%%%%%%%%%%%%%%%

\begin{figure}[tb]
\centering{%
\includegraphics[width=0.99\columnwidth,clip, trim=15 10 15 10]{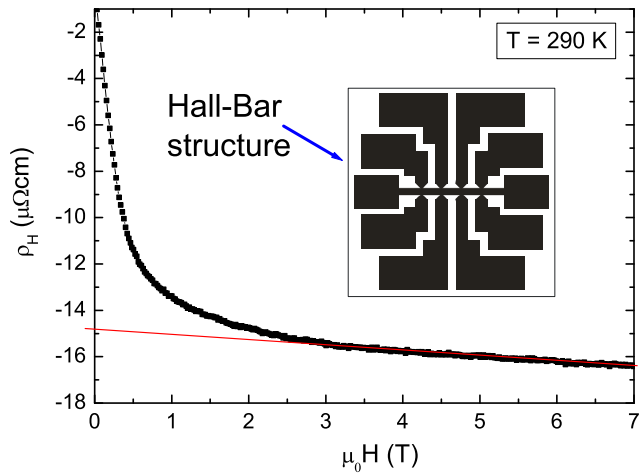}}
 \caption{Hall effect for a Fe$_3$O$_4$ thin film grown on (001) MgO. The dotted line
is a fit to the high field data, where the film magnetization saturates. The
inset shows an optical micrograph of the Hall bar geometry.}
 \label{Reisinger_Fig3}
\end{figure}

In Fig.~\ref{Reisinger_Fig3}, the Hall resistivity is plotted versus the
applied magnetic field at $T=290$\,K. It is well known that in nonmagnetic
metals the familiar Hall current arises when electrons moving in crossed
electric ($\mathbf{E}$) and magnetic fields ($\mathbf{H}$) are deflected by the
Lorentz force. However, in a ferromagnet subject to $\mathbf{E}$ alone, an
anomalous Hall current appears transverse to $\mathbf{E}$. Karplus and
Luttinger\cite{Karplus:1954a} proposed a quantum mechanical origin of the Hall
current, where an electron in the conduction band spends part of its time in
nearby bands thereby acquiring a spin-dependent anomalous velocity. In modern
terms, this anomalous velocity is related to the Berry phase and recently has
been applied to explain the anomalous Hall effect (AHE) in Mn-doped
GaAs\cite{Jungwirth:02}. A more conventional explanation is based on skew
scattering and side jump\cite{Noziere:1973a}. Unfortunately, the anomalous Hall
effect in itinerant ferromagnets is still discussed controversially. In an AHE
experiment the observed Hall resistivity $\rho_H$ comprises two terms,
\begin{equation}
\rho_H = R_0 \mu_0 H + R_A \mu_0 M \; ,
\end{equation}
where $R_0$ is the ordinary and $R_A$ the anomalous Hall
coefficient. It is evident that after the magnetization $M$ has
saturated on increasing the applied field $H$, the {\em change} of
$\rho_H$ is only due to the normal Hall effect. From the linear
high field dependence of the Hall effect data we derive $R_0
\simeq -2.12\cdot10^{-9}$\,m$^3$/C at 300\,K. With the relation
$R_0=1/en$ we obtain the electron density
$n=2.95\cdot10^{21}$/cm$^3$. With the density
$\rho=5.185$\,g/cm$^3$ we then obtain the number of electrons per
Fe$_3$O$_4$ formula unit to 0.22. This value is comparable to that
observed in bulk single crystal samples\cite{Lavine:59}. The
negative sign of the ordinary Hall coefficient suggests electron
conduction. We also note that $R_0$ is slightly larger at 160\,K
as compared to 300\,K due to the increase in the Hall mobility.
The Hall mobility $\mu_H$ is given by the product of the ordinary
Hall coefficient and the conductivity. At 300\,K we obtain $\mu_H
\simeq 0.48$\,cm$^2$/Vs. The anomalous Hall coefficient at 300\,K
is determined to $R_A \simeq -2.5\cdot10^{-7}$\,m$^3$/As using the
measured saturation magnetization of 3.6\,$\mu_B$ per formula
unit. We note that in the half-metallic double perovskite
Sr$_2$FeMoO$_6$ $R_A$ shows a positive sign, while $R_0$ is
negative\cite{Westerburg:00}. This is in contrast to magnetite,
where both $R_0$ and $R_A$ have negative sign. Within a simple
model the different signs of $R_A$ for different materials can be
understood in terms of asymmetric scattering due to different
densities of states available for positive and negative orbital
orientations\cite{Fert:1972}.

%%%%%%%%%%%%%%%%%%%%% FIGURE 3: Hall Effect %%%%%%%%%%%%%%%%%%%%%%%%%%
%%%%%%%%%%%%%%%%%%%%%%%%%%%%%%%%%%%%%%%%%%%%%%%%%%%%%%%%%%%%%%%%%%%%%%%

Discussing $\sigma$ and $n$ we can state that in a simple physical picture one
would expect that the minority spin electron that is shared by the two
octahedrally coordinated B-sites gives rise to electronic conductivity above
the Verwey transition. Below the Verwey transition, charge ordering of the
B-site ions leads to an insulating state. However, the physical picture of
charge ordering below the Verwey transition has been challenged for example by
resonant x-ray scattering measurements\cite{Garcia:01}. The observation that
only 0.22 electrons per f.u.~contribute to the conductivity could be explained
by localization effects due to electronic correlations. Recently, in an {\em ab
initio} study of charge order in magnetite the local spin density has been
calculated as a function of symmetry (or structural distortion). This
calculation yields 100\% spin polarization, a total spin magnetic moment of
about 4.0\,$\mu_B$ and about 0.24 electronic states per formula unit at the
Fermi level\cite{Szotek:03}. Although these values have been calculated for the
charge ordered ground state, they are remarkably consistent with the results
obtained in our study above the Verwey transition.

In summary, we have grown high quality epitaxial thin films of magnetite with
properties comparable to those of best bulk samples. For the first time, we
report Hall effect measurements for thin films, indicating that only 0.22
electrons per formula unit contribute to electrical conductivity.

This work was supported by the Deutsche Forschungsgemeinschaft (Al/560) and the
Bundesministerium f\"{u}r Bildung und Forschung (project 13N8279).

\end{document}